\documentclass[%
reprint,
 amsmath,amssymb,
 aps,
]{revtex4-1}
\usepackage{graphicx}
\usepackage{braket}
\usepackage{bm}
\usepackage{amsmath}
\usepackage{ulem}
\begin{document}
\title{4$\alpha$ linear-chain state produced by $^{9}$Be+$^{9}$Be collision}
\author{Tomoyuki Baba$^1$, Yasutaka Taniguchi$^{2,6}$ Masaaki Kimura$^{3,4,5,6}$}
\affiliation{$^1$Kitami Institute of Technology, 090-8507 Kitami, Japan\\
$^2$Department of Information Engineering, National Institute of Technology (KOSEN), Kagawa College, 769-1192 Mitoyo, Japan\\
$^3$Department of Physics, Hokkaido University, 060-0810 Sapporo, Japan\\
$^4$RIKEN Nishina Center, 351-0198 Wako, Japan.\\
$^5$Reaction Nuclear Data Centre (JCPRG),  Hokkaido University, 060-0810 Sapporo, Japan\\
$^6$Research Center for Nuclear Physics (RCNP), Osaka University, 567-0047 Ibaraki, Japan}
\date{\today}

\begin{abstract}
Extreme nuclear deformations provide great insight into the geometric formation of quantum many-body systems.
In this work, the $4\alpha+2n$ linear chain is assessed in $^{18}$O.
We predict excitation energies, moment-of-inertia, $\alpha$-, and $^{9}$Be-decay widths by using the antisymmetrized molecular dynamics.
We show that the $K^\pi=3^-$ linear-chain states may be verified by the head-on $^{9}{\rm Be}+{}^{9}{\rm Be}$ collision experiments.
\end{abstract}

\maketitle

{\it Introduction}---Nuclear clustering has many similarities to molecules.
For instance, extra neutrons surrounding $\alpha$-particles play a glue-like role analogous to the covalent electrons of atomic molecules as in beryllium ($2\alpha+xn$) \cite{seya81,oertzen96,enyo99,12Beex1,itagaki00,itoBe,ito12Be}, carbon ($3\alpha+xn$) \cite{itag01}, and in the heavier mass isotopes \cite{furu08,kimu07,tani14}.
In the carbon isotopes ($3\alpha+xn$), it has been expected that the extremely deformed states which have the intrinsic structure of linearly aligned three $\alpha$-particles, called as linear-chain (LC) structure, will be stabilized by the assist of the glue-like role of the covalent neutrons.
Recent experimental \cite{free14,frit16,tian16,yama17,li17} and theoretical \cite{suha10,ebra14,baba16,baba17,enyo20,ren20} studies have identified the linear-chain states in $^{14}$C($3\alpha+2n$), and the research of $^{16}$C ($3\alpha+4n$) has also been making progress \cite{baba18,liu20,baba20}.

A fascinating and long-standing open question is how many $\alpha$-particles can compose the linear-chain structure.
Unlike the 3$\alpha$ linear chains, linear chains containing four or more $\alpha$-particles have not been identified.
In order to produce the $4\alpha$ linear-chain state in oxygen isotopes, the $^{A}$Be+$^{A}$Be resonant scattering may be a natural way.
Although the formation of the $4\alpha$ linear chain in $^{16}$O was predicted by several theoretical works \cite{suzu72, ichi11,suha14,yao14,suzu17,inak18},
$^{8}$Be is the unbound nucleus, so that it is not easy to prove its existence from the $^{8}$Be+$^{8}$Be reaction.
Given that $^{9}$Be is the only stable beryllium isotope, $^{9}$Be+$^{9}$Be scattering is the most feasible way to confirm the $4\alpha+2n$ linear chain.
Therefore, we study the excited states of $^{18}$O to search for the candidate of $4\alpha+2n$ linear chain.

\begin{figure}[h]
 \centering
 \includegraphics[width=1.0\hsize]{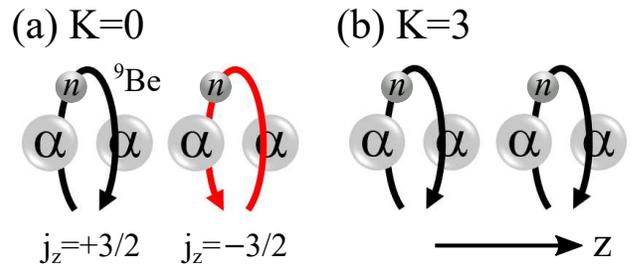}
 \caption{(color online) Schematic illustration of the $4\alpha$ linear chains constructed by (a) anti-parallel and (b) parallel alignments with respect to valence neutron's $j_z$ of $^{9}$Be.
 The black arrows indicate $j_z=+3/2$ orbit of the valence neutron while the red arrow indicates $j_z=-3/2$. } 
 \label{fig:beBrink}
\end{figure}

{\it Linear chain produced by $^{9}$Be+$^{9}$Be collision}---Let us consider the $4\alpha$ linear-chain configurations which will be produced by the head-on $^{9}$Be+$^{9}$Be collisions.
For simplicity, we approximate the ground state of $^9$Be ($3/2^-$) as the $\alpha+\alpha+n$ system with a valence neutron occupying the $p_{3/2}$ orbit with $j_z=\pm3/2$.
As illustrated in Fig. \ref{fig:beBrink}, there are two ways to linearly align two $^9$Be, which yields different valence neutron configurations; (a) the anti-parallel and (b) the parallel alignments with respect to the valence neutron's $j_z$.
The anti-parallel alignment yields the intrinsic state with $K=0$, where $K$ denotes the $z$-component of the intrinsic angular momentum equal to the sum of valence neutron's $j_z$.
Because this configuration is an admixture of the positive- and negative-parity states, we expect that it leads to a pair of the rotational bands; $J^\pi=0^+, 2^+, 4^+, ...$ and $1^-, 3^-, 5^-, ...$ bands.
Indeed, we obtain both bands in this work.
However, we focus on the only $J^\pi=0^+, 2^+, 4^+, ...$ band because the negative-parity band is located at higher than the positive-parity band.

The parallel alignment (Fig. \ref{fig:beBrink} (b)) yields the intrinsic state with $K=3$.
The parity of this configuration is uniquely determined as negative due to the following reason.
The spin wave function is symmetric ($S=1$) as both neutron spins are aligned to $s_z=1/2$.
Because the isospin wave function is also symmetric, the spatial wave function must be asymmetric with respect to the exchange of two valence neutrons (parity transformation).
Thus, this configuration forms a $J^\pi=3^-, 4^-, 5^-, ...$ band.
In short, the anti-parallel alignment yields a pair of the positive- and negative-parity bands with $K=0$, whereas the parallel alignment yields a negative-parity band with $K=3$.

{\it Calculated properties of the linear-chain states}---In order to describe the linear-chain states, we use the antisymmetrized molecular dynamics (AMD).
We employ the Hamiltonian with the Gogny D1S nucleon-nucleon interaction \cite{gogn91}.
The AMD wave function $\Psi^{\pi}_{\rm AMD}$ is a parity-projected Slater determinant of single particle wave packets, 
\begin{align}
 \Psi^{\pi}_{\rm AMD} =\hat{P}^{\pi}\Psi_{\rm AMD} =\hat{P}^{\pi}{\mathcal A} \{\psi_1,\psi_2,...,\psi_A \}.
 \label{eq:amdwf}  
\end{align}
Here, $\hat{P}^{\pi}$ is the parity-projection operator, and $\psi_i$ is the single particle wave packet which is a direct product of the deformed
Gaussian for the spatial part, spin ($\chi_i$) and isospin ($\xi_i$) parts \cite{kimu04},  
\begin{align}
 \phi_i({\bm r}) &= \prod_{\sigma=x,y,z}\exp\biggl\{-\nu_\sigma\Bigl(r_\sigma -\frac{Z_{i\sigma}}{\sqrt{\nu_\sigma}}\Bigr)^2\biggr\}\otimes \chi_i\otimes \xi_i, \\
 \chi_i &= a_i\chi_\uparrow + b_i\chi_\downarrow,\quad
 \xi_i = {\rm proton} \quad {\rm or} \quad {\rm neutron}.\nonumber
\end{align}
The centroids of the Gaussian wave packets $\bm Z_i$, the direction of nucleon spin $a_i, b_i$,
and the width parameter $\nu_\sigma$ are the variables determined by the frictional cooling method \cite{enyo95}.
In this study, we impose the constraint on the quadrupole deformation parameter $\beta$ to describe extremely deformed $4\alpha$ linear chain.
After the variational calculation, the eigenstate of the total angular momentum $J$ is projected out.
We perform the generator coordinate method \cite{hill54} by employing the quadrupole deformation parameter $\beta$ as the generator coordinate.

In our previous work \cite{baba19}, it has been shown that the AMD plausibly describes the low-lying states of $^{18}$O.
The binding energy of $^{18}$O is calculated as 139.97 MeV whereas the observed value is 139.81 MeV.
The low-lying excited states including the $^{14}{\rm C}+{}^{4}{\rm He}$ cluster states are also reasonably described.
Therefore, we expect that the AMD can also precisely describe higher-lying states of $^{18}$O.
\begin{figure}[h]
 \centering
 \includegraphics[width=1.0\hsize]{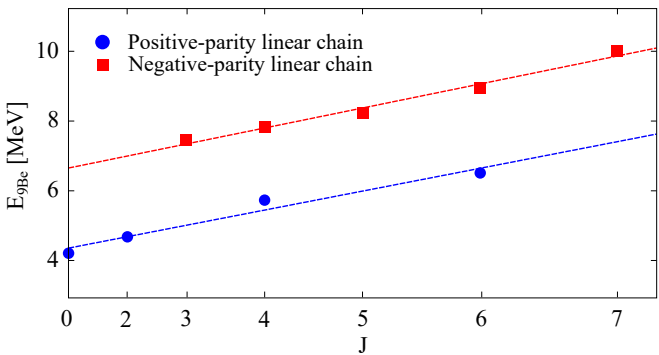}
 \caption{(color online) Calculated energies above $^{9}{\rm Be}+{}^{9}{\rm Be}$ threshold of the linear-chain states as a function of angular momenta.
 The energy is relative to 23.64 MeV.} 
 \label{fig:spec}
\end{figure}

\begin{figure}[h]
 \centering
 \includegraphics[width=1.0\hsize]{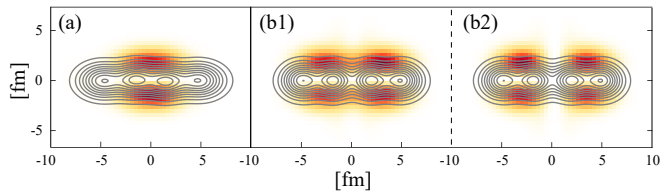}
 \caption{(color online) Density distributions of the intrinsic states of the linear-chain bands.
 Contour lines show the proton density distributions and color plots show the valence neutron orbits.
 Panel (a) shows the intrinsic state of the positive-parity band, in which two valence neutrons occupy the negative-parity orbits with $j_z=\pm3/2$.
 Panel (b1) and (b2) show the intrinsic state of the negative-parity band, in which a valence neutron occupies the negative-parity orbit with $j_z=3/2$ (b1),
 and the other occupies the positive-parity orbit with $j_z=3/2$ (b2).} 
 \label{fig:dens}
\end{figure}

Note that the present calculation does not assume a prior linear-chain configuration.
In fact, we obtain many excited states with various cluster and non-cluster states.
Among these excited states, we assign two rational bands as the linear-chain candidates shown in Fig. \ref{fig:spec}; a positive-parity band built on the $0^+$ state followed by $2^+, 4^+, ...$ states
and a negative-parity band built on the $3^-$ state followed by $4^-, 5^-, ...$ states.
We remark that the spin-parity of these bands is consistent with that expected from the $^{9}{\rm Be}+{}^{9}{\rm Be}$ configurations discussed above.
The reasons for this assignment are as follows.
The member states of each band have the same intrinsic structure.
All positive-parity band members have a large overlap with the intrinsic wave function shown in Fig. \ref{fig:dens}(a).
Its proton density distribution shows the linear alignment of four $\alpha$ particles, and the two valence neutrons occupy the negative-parity orbits with $j_z=\pm3/2$.
This intrinsic structure is approximately the anti-parallel configuration shown in Fig. \ref{fig:beBrink}(a).
However, differently from Fig. \ref{fig:beBrink}(a), the two valence neutrons are localized around two $\alpha$-particles at the center because of the attraction between valence neutrons and no Pauli exclusion.
Therefore, the total system forms the $\alpha+{}^{10}{\rm Be}+\alpha$ -like structure.

The intrinsic state of the negative-parity band is shown in Fig. \ref{fig:dens} (b1) and (b2).
The proton density distribution indicates that this band also has 4$\alpha$ linear-chain core.
A valence neutron occupy the negative-parity orbit with $j_z=3/2$ (Fig. \ref{fig:dens}(b1)),
and the other occupies the positive-parity orbit with $j_z=3/2$ (Fig. \ref{fig:dens}(b2)).
These single-particle orbits can be understood as a linear combination of the $p_{3/2}$ orbits of two $^9$Be.
Let us denote the $p_{3/2}$ ($j_z=3/2$) orbit of the left (right) side $^9$Be as $\varphi_L$ ($\varphi_R$), which are schematically illustrated as black arrows in Fig. \ref{fig:beBrink}(b).
Then, the single-particle orbits are represented as,
\begin{align}
 \varphi_\pm = \frac{1}{\sqrt{2}}(\varphi_L \pm \varphi_R). \nonumber
\end{align}
They generate an orthogonalized pair of the negative- and positive-parity orbits with $j_z=3/2$, which corresponds to Fig. \ref{fig:dens} (b1) and (b2), respectively.
This intrinsic state corresponds to the parallel configuration in Fig. \ref{fig:beBrink}.
In contrast to the anti-parallel configuration, the two valence neutrons separately locate the left and right side Be because of the Pauli principle which results from their parallel spin.
We find that only these two bands have the structure corresponding to the $^{9}{\rm Be}+{}^{9}{\rm Be}$ head-on collision in the vicinity of the energies near the $^{9}{\rm Be}+{}^{9}{\rm Be}$ threshold.

These two bands have strongly deformed intrinsic shapes compatible with the $4\alpha$ linear chain.
The quadrupole deformation parameters of the positive- and negative-parity intrinsic states are equally $\beta=1.34$.
Consequently, they have enormous moment-of-inertia as large as $\hbar/2\Im=55$ keV for the positive-parity band and $58$ keV for the negative-parity band.
These are even larger than a rigid-rotor estimation, $\hbar/2\Im=85$ keV estimated as follows.
The classical moment-of-inertia of the prolate spheroid with 2:1, 3:1, and 4:1 deformation satisfy the following relation,
\begin{align}
\Im_{2:1}:\Im_{3:1}:\Im_{4:1}=5:15:34. \nonumber
\end{align}
Since the moment-of-inertia of the $2\alpha$ and $3\alpha$ linear chains are measured as $\hbar/2\Im=590$ keV ($2\alpha+2n$) \cite{till04} and $\hbar/2\Im=190$ keV in ($3\alpha+2n$) \cite{yama17},
the moment-of-inertia of $4\alpha$ linear chain is estimated as $\hbar/2\Im=85$ keV.
\begin{table*}[h!tb]
 \caption{Excitation energies, energies above $^{9}{\rm Be}+{}^{9}{\rm Be}$ threshold, partial $\alpha$-decay widths and dimensionless reduced widths of the $4\alpha$ linear-chain states.
 $^{14}$C is assumed to be the $0^+$ and $2^+$ states of the linear-chain band. The channel radius is $7.0$ fm.}
\label{tab:widtha}
\begin{center}
 \begin{ruledtabular}
  \begin{tabular}{lcccccl} 
    $J^\pi$ & $E_x$ [MeV] & $E_{{\rm Be}}$ [MeV] & $\Gamma_\alpha(0^+$; LC) [keV] & $\theta^2_{\alpha}(0^+$; LC) $[\times10^{-2}]$ & $\Gamma_\alpha(2^+$; LC) [keV] & $\theta^2_{\alpha}(2^+$; LC) $[\times10^{-2}]$ \\
   \hline
      $0^+$ & 27.85 & 4.21 & 382 & 8.21 & 1136 & 31.8   \\
      $2^+$ & 28.32 & 4.68 & 346 & 7.70 & 1044 & 13.4 ($l=4$)  \\
      $4^+$ & 29.37 & 5.73 & 288 & 3.51 & 918 & 5.37 ($l=2$)   \\
      $6^+$ & 30.16 & 6.51 & 160 & 5.39 & 572 & 10.2 ($l=4$)   \\
      $3^-$ & 31.09 & 7.45 & 0 & 0.00 & 0 & 0.00  \\
      $4^-$ & 31.46 & 7.82 & 0 & 0.00 & 0 & 0.00  \\
      $5^-$ & 31.86 & 8.22 & 0 & 0.00 & 0 & 0.00  \\
      $6^-$ & 32.58 & 8.95 & 0 & 0.00 & 0 & 0.00  \\
      $7^-$ & 33.68 & 10.0 & 0 & 0.00 & 0 & 0.00  \\
  \end{tabular}
 \end{ruledtabular}
 \end{center}
\end{table*}

{\it Decay properties of the linear-chain states}---In order to take a deep dive on the difference of the two linear-chain bands, we discuss their decay properties.
Calculated excitation energies, partial $\alpha$-decay widths, and dimensionless reduced widths are listed in Table. \ref{tab:widtha}.
A dimensionless reduced width $\theta^2_{l}(a)$ is defined by the ratio of the reduced width to Wigner limit,
\begin{align}
 \theta^2_{l}(a) = \frac{a}{3}|ay_{l}(a)|^2,
\end{align}
where $y_{l}(a)$ is the reduced width amplitude,
\begin{align}
 y_{l}(r) = \sqrt{\frac{A!}{A_{\rm C1}!A_{\rm C2}!}}
 \langle \phi_{\rm C1}[\phi_{\rm C2}Y_{l0}({\hat r})]_{J^\pi M}
 |\Psi^{J\pi}_{Mn}\rangle,\label{eq:rwa}
\end{align}
In the $\alpha+{}^{14}{\rm C}$ channel, the daughter nucleus $^{14}$C is assumed to be the $0^+$ and $2^+$ states of the $3\alpha$ linear-chain states of $^{14}$C studied in Ref.\cite{baba16}.
The reduced width in the $\alpha+{}^{14}{\rm C}$(g.s.) is almost zero due to the extreme deformation of the $4\alpha$ linear chain.
The $\alpha$-decay properties are different between the $K^\pi=0^+$ and $K^\pi=3^-$ linear chains.
The $K^\pi=0^+$ linear-chain states have very large $\alpha$-decay widths into the linear-chain states of $^{14}$C.
In particular, the widths to the $\alpha+{}^{14}{\rm C}(2^+$; LC) channel are huge because of the strong angular correlation between the linearly aligned $\alpha$-particles.
This character is in contrast to the Hoyle state where $\alpha$ particles are weakly bound with $l=0$ and hence, the $^{8}{\rm Be}(0^+_1)$ component dominates \cite{funa15}. 
The large reduced width amplitude for $J^\pi= 0^+, 2^+, ...$ is a significant linear-chain structure feature.
As the $^{14}$C orientation is fixed, it does not become an eigenstate but a mixed state of angular momentum.
The similar character was also discussed for the $3\alpha$ linear chain of the carbon isotopes \cite{suzu72,baba16,baba18}.
The $K^\pi=3^-$ linear-chain states have almost zero $\alpha$-decay widths into the linear chain of $^{14}$C.
It is clear because the both valence neutrons in the linear-chain states of $^{14}$C are negative parity, as a result they are orthogonal with the linear chain of $^{18}$O with the positive- and negative-parity valence neutrons shown in Fig. \ref{fig:dens} (b).
In addition, we did not find the negative-parity $3\alpha$ linear-chain states of $^{14}$C with the positive- and negative-parity valence neutrons \cite{baba16}.
In present calculation, therefore, the $K^\pi=3^-$ linear-chain states do not decay to any $\alpha+{}^{14}{\rm C}$ channels.
\begin{table*}[h!tb]
 \caption{Partial $^{9}{\rm Be}$-decay widths and dimensionless reduced widths of the $4\alpha$ linear-chain states.
 $^{9}{\rm Be}(3/2^-)+{}^{9}{\rm Be}(3/2^-)$ and $^{9}{\rm Be}(3/2^-)+{}^{9}{\rm Be}(5/2^-)$ are assumed. The channel radius is $7.0$ fm.}
\label{tab:widthb}
\begin{center}
 \begin{ruledtabular}
  \begin{tabular}{lccclcc} 
    $J^\pi$ & $E_x$ [MeV] & $E_{{\rm Be}}$ [MeV] & $\Gamma_{\rm Be}(3/2^-)$ [keV] & $\theta^2_{\rm Be}(3/2^-) [\times10^{-2}]$ & $\Gamma_{\rm Be}(5/2^-)$ [keV] & $\theta^2_{\rm Be}(5/2^-) [\times10^{-2}]$ \\
   \hline
      $0^+$ & 27.85 & 4.21 & 6 & 0.31 & 9 & 0.52 ($l=2$)  \\
      $2^+$ & 28.32 & 4.68 & 6 & 0.30 & 8 & 0.23 ($l=4$)  \\
      $4^+$ & 29.37 & 5.73 & 6 & 0.14 & 8 & 0.10 ($l=2$)  \\
      $6^+$ & 30.16 & 6.51 & 4 & 0.32 & 7 & 0.25 ($l=4$)  \\
      $3^-$ & 31.09 & 7.45 & 264 & 5.37 ($l=1$) & 253 & 1.84 ($l=1$) \\
      $4^-$ & 31.46 & 7.82 & 151 & 3.02 ($l=3$) & 144 & 0.92 ($l=3$) \\
      $5^-$ & 31.86 & 8.22 & 335 & 5.81 ($l=3$) & 278 & 2.21 ($l=3$) \\
      $6^-$ & 32.58 & 8.95 & 307 & 6.33 ($l=5$) & 245 & 2.10 ($l=5$) \\
      $7^-$ & 33.68 & 10.0 & 281 & 4.87 ($l=5$) & 251 & 1.51 ($l=5$) \\
  \end{tabular}
 \end{ruledtabular}
 \end{center}
\end{table*}

Partial $^{9}{\rm Be}$-decay widths, and dimensionless reduced widths are listed in Table. \ref{tab:widthb}.
We consider the $^{9}{\rm Be}(3/2^-)+{}^{9}{\rm Be}(3/2^-)$, $^{9}{\rm Be}(3/2^-)+{}^{9}{\rm Be}(5/2^-)$ channels, where ${}^{9}{\rm Be}(3/2^-)$ and ${}^{9}{\rm Be}(5/2^-)$ are the ground and excited states of ${}^{9}{\rm Be}$.
Characters of the decay widths are different between the $K^\pi=0^+$ and $K^\pi=3^-$ linear chains.
The $K^\pi=0^+$ linear-chain states have very small widths into the $^{9}{\rm Be}+{}^{9}{\rm Be}$ channel, which is consistent with the $\alpha+{}^{10}{\rm Be}+\alpha$ picture shown in Fig. \ref{fig:dens} (a).
On the other hand, the $K^\pi=3^-$ linear-chain states show large $^{9}{\rm Be}$ widths.
Therefore, the $K^\pi=3^-$ linear chain manifests so strong $^{9}$Be+$^{9}$Be correlation shown in Fig. \ref{fig:beBrink} (b) and these linear-chain states can be observed by a $^{9}$Be+$^{9}$Be collision.

\begin{figure}[h]
 \centering
 \includegraphics[width=1.0\hsize]{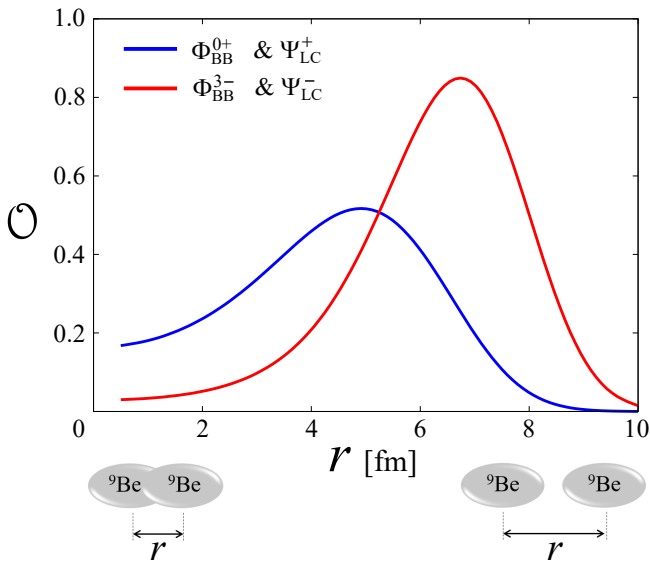}
 \caption{(color online) Calculated overlaps between the $^{9}$Be+$^{9}$Be Brink wave functions and the AMD wave functions. $\Psi^+_{LC}$ and $\Psi^-_{LC}$ correspond to Fig. \ref{fig:dens} (a) and (b), respectively.}
 \label{fig:beovlp}
\end{figure}

In order to clarify the characteristic $^{9}{\rm Be}$-decay modes, we calculate overlaps between the Brink and the AMD wave functions defined as,
\begin{align}
 \mathcal{O}(r) = \frac{|\langle \Phi^{K\pi}_{\rm BB}(r)|\hat{P}^J_{KK}|\Psi^\pi_{\rm AMD}\rangle|^2}{|\langle \Phi^{K\pi}_{\rm BB}(r)|\hat{P}^J_{KK}|\Phi^{K\pi}_{\rm BB}(r)\rangle||\langle \Psi^\pi_{\rm AMD}|\hat{P}^J_{KK}|\Psi^\pi_{\rm AMD}\rangle|}, \label{eq:ovlp}
\end{align}
where $\hat{P}^J_{KK}$ is the angular momentum projection operator.
The Brink wave functions $\Phi^{K\pi}_{\rm BB}(r)$ are constructed by the linearly alignment of two $^{9}$Be shown in Fig. \ref{fig:beBrink}; (a) anti-parallel $\Phi^{0+}_{\rm BB}(r)$ and (b) parallel $\Phi^{3-}_{\rm BB}(r)$ alignments.
\begin{align}
 \Phi^{0+}_{\rm BB}(r) &= \hat{P}^{\pi}\mathcal{A} \{ \phi^{jz=3/2}_{\rm Be}(-r/2)\phi^{jz=-3/2}_{\rm Be}(r/2) \}, \label{eq:oppo}\\
 \Phi^{3-}_{\rm BB}(r) &= \hat{P}^{\pi}\mathcal{A} \{ \phi^{jz=3/2}_{\rm Be}(-r/2)\phi^{jz= 3/2}_{\rm Be}(r/2) \}. \label{eq:same}
\end{align}
Here, the wave function of $^{9}$Be is described as,  
\begin{align}
 \phi^{jz}_{\rm Be} &= \mathcal{A} \{ \phi_\alpha \otimes \phi_\alpha \otimes (0p_{3/2}) \}, \label{eq:be9}
\end{align}
where $(0p_{3/2})$ is represented by a infinitesimally shifted Gaussian wave packet based on the antisymmetrized quasicluster model \cite{suhajj}.
Figure \ref{fig:beovlp} shows the calculated overlap as a function of the distance $r$.
The $K^\pi=3^-$ linear chain has a large amount of overlap 0.84 with the Brink wave function $\Phi^{3-}_{\rm BB}$($r=6.5$ fm) shown in Fig. \ref{fig:beBrink} (b).
On the other hand, the $K^\pi=0^+$ linear chain has a amount of overlap 0.52 with $\Phi^{0+}_{\rm BB}$($r=5.0$ fm) shown in Fig. \ref{fig:beBrink} (a).
Compared to the $K^\pi=3^-$ linear chain, the $^{9}$Be+$^{9}$Be correlation is small in the $K^\pi=0^+$ linear chain.
The $K^\pi=3^-$ and $K^\pi=0^+$ states have different features in the outer region. The $K^\pi=3^-$ state has more considerable overlap in the outer region than the $K^\pi=0^+$ state, with the outer peak position.
At the present channel radius of $7.0$ fm, the overlap of $K^\pi=3^-$ is one order of magnitude larger than that of $K^\pi=0^+$, which leads to the much larger $^{9}$Be+$^{9}$Be decay width.
Indeed, this difference is reflected in the difference of the $^{9}$Be-reduced widths in Table. \ref{tab:widthb}.
The $K^\pi=0^+$ linear chain has the weak $^{9}$Be+$^{9}$Be correlation so that the $^{9}$Be-reduced widths are small,
while the $K^\pi=3^-$ linear chain has so strong $^{9}$Be+$^{9}$Be correlation that the member states have large $^{9}$Be-reduced widths.
Therefore, we conclude that the $K=3^-$ linear-chain states of $^{18}$O can be observed by a $^{9}$Be+$^{9}$Be resonant scattering shown in Fig. \ref{fig:beBrink} (b).

{\it Summary}---In summary, we have presented the first assessment of the $4\alpha+2n$ linear-chain configuration in $^{18}$O using the AMD calculation.
There are two different $4\alpha$ linear-chain bands, $K^\pi=0^+$ and $K^\pi=3^-$. 
We predict their excitation energies, moment-of-inertia, $\alpha$-, and $^{9}$Be-decay widths.
In both bands, the moment-of-inertia is rather large, which is a strong evidence for the extreme deformation.
The two types of linear chains show different decay properties.
The $K^\pi=0^+$ linear-chain states show large decay widths into the $\alpha+{}^{14}{\rm C}$(LC) channel while the $K^\pi=3^-$ linear-chain states show large those into the $^{9}{\rm Be}+{}^{9}{\rm Be}$ channel.
In order to clarify this difference, we calculate the overlaps with the $^{9}$Be+$^{9}$Be Brink wave functions.
As the results, the $K^\pi=0^+$ linear chain has small overlap while the $K^\pi=3^-$ linear chain has large overlap.
It means that the $K^\pi=3^-$ linear chain shows a strong $^{9}$Be+$^{9}$Be correlation.
Therefore, we expect that the linear chain of $^{18}$O can be produced by the head-on $^{9}$Be+$^{9}$Be collision.
We believe that these are promising properties that can be investigated in future experiments and establish the existence of the exotic $4\alpha$ linear chain.

We thank Prof. M. Ito and Dr. H. Yamaguchi for useful discussions.
This calculation has been done on a supercomputer at Research Center for Nuclear Physics, Osaka University.
The authors acknowledge the support of the collaborative research program 2021 at Hokkaido University.
This work was supported by the JSPS KAKENHI Grant No. 19K03859.

\end{document}